\documentclass[aps,pra,superscriptaddress,preprint,amsmath,amssymb,showpacs]{revtex4-2}
\usepackage[english]{babel}
\usepackage{amssymb,bm}
\usepackage{graphicx}
\usepackage{amsmath}
\usepackage{color}
\usepackage{comment}
\usepackage{subcaption}
\usepackage[colorlinks=true,citecolor=blue,urlcolor=blue]{hyperref}
\begin{document}

\begin{titlepage}
\title{Charge asymmetry in the process $e^+e^-\to \pi^+\pi^-\pi^0\gamma$}
\author{I.V. Obraztsov}\email{ivanqwicliv2@gmail.com}
\author{A.S. Rudenko}\email{a.s.rudenko@inp.nsk.su}
\author{A.I. Milstein}\email{a.i.milstein@inp.nsk.su}
\affiliation{Budker Institute of Nuclear Physics of SB RAS, 630090 Novosibirsk, Russia}
\affiliation{Novosibirsk State University, 630090 Novosibirsk, Russia}

\date{\today}

\begin{abstract}
	The charge asymmetry in the process $e^+e^-\to \pi^+\pi^-\pi^0\gamma$ is studied for photons with energies $\nu\ll\sqrt{s}$.  The longitudinal polarization of electrons (positrons) is taken into account. The asymmetry arises due to  interference of the  amplitudes for production of pion states with opposite  $C$ parities. The contribution of $a_1^0(1260)$ meson in the intermediate state  to the charge asymmetry is discussed.
	This contribution is not the leading one, however it is not negligible. It is shown that the charge asymmetry can reach several tens of percent.
	\end{abstract}

\maketitle
\end{titlepage}
\section{Introduction}
At present, to describe the processes of strong interactions at large distances $r\gg 1/\Lambda_{QCD}$, it is necessary to use phenomenological models.
Direct production of mesons with the quantum numbers $J^{PC}=1^{++}$ in $e^+e^-$ annihilation is of great interest since such particles have positive $C$-parity and can be produced as a result of the fusion of two virtual photons. Since  a particle with spin one cannot decay into two real photons (Landau-Yang theorem~\cite{Landau:48, Yang:50}),  direct production of mesons with quantum numbers $J^{PC}=1^{++}$ allows one to investigate in detail their internal structure describing by the corresponding form factors.

An example of  meson with quantum numbers $J^{PC}=1^{++}$ is the isoscalar $f_1(1285)$. The production of $f_1(1285)$ meson by two virtual photons in the process $e^+e^-\to 4\pi$ was  observed in the experiment~\cite{f1exp2020}. The experimental results are consistent with the theoretical predictions in Ref.~\cite{f1th2020}. Another example of a particle with the quantum numbers $J^{PC}=1^{++}$ is the isovector $a_1(1260)$. The leading contribution to the width of $a_1(1260)$ is given by the decay mode  $a_1(1260)\to 3\pi$ \cite{ParticleDataGroup:2024cfk}. Therefore, it is possible to observe the direct production of $a^0_1(1260)$ by two virtual photons in the process $e^+e^-\to \pi^+\pi^-\pi^0$. Note that the main contribution to the cross section of this process is made by vector mesons ($\omega(782)$, $\phi(1020)$, etc.), which are produced through one virtual photon, since the amplitude $e^+e^-\to a^0_1(1260)$ contains additional suppression by the fine structure constant $\alpha$.
 
 The one-photon and two-photon amplitudes in the process $e^+e^-\to \pi^+\pi^-\pi^0$ have different $C$-parities, so that there is a nonzero charge asymmetry proportional to the interference of these amplitudes. Therefore, it is easier to observe the direct production of $a^0_1(1260)$ via two virtual photons by measuring the charge asymmetry rather than the total cross section. Besides $a^0_1(1260)$ meson,  $a_2^0(1320)$ meson can also be produced via two virtual photons and decay into $\pi^+\pi^-\pi^0$. However, its contribution to the cross section is noticeably less than that of $a^0_1(1260)$, and we do not discuss  $a_2^0(1320)$ in our work.  

The $a_1(1260)$ meson was studied experimentally in hadronic reactions \cite{Amsterdam-CERN-Nijmegen-Oxford:1977eoy,ACCMOR:1980llh,Longacre:1981ax,Dankowych:1981ks,Zielinski:1984au,WA76:1990utn,Ando:1990ti,WA102:1998inv,WA102:2001rbg,Chung:2002pu,Belle:2002gzj,OBELIX:2004oio,FOCUS:2007ern,COMPASS:2009xrl,dArgent:2017gzv,LHCb:2017swu,COMPASS:2018uzl,COMPASS:2020yhb}, in decays of $\tau$ lepton \cite{DELCO:1986bqg,ARGUS:1986yqj,Schmidke:1986gp,Tornqvist:1987ch,MAC:1987pno,Isgur:1988vm,Bowler:1988kf,Ivanov:1989qw,ARGUS:1992olh,OPAL:1997was,DELPHI:1998bhv,CLEO:1999rzk,ALEPH:1999uux,Bondar:1999ac,GomezDumm:2003ku,CLEO:2004hrb,JPAC:2018zwp}, in the processes $e^+e^-\to a_1(1260)\pi\to 4\pi$ \cite{CMD-2:1998gab} and $e^+e^-\to a_1(1260)2\pi\to 5\pi$ \cite{BaBar:2007qju}. However, the form factor at the vertex $\gamma^*\gamma^*a^0_1(1260)$ has not yet been experimentally studied. To investigate this form factor, it is possible to  study the charge asymmetry in the processes $e^+e^-\to \pi^+\pi^-\pi^0$ and $e^+e^-\to \pi^+\pi^-\pi^0\gamma$. The first one was discussed in our  work \cite{ORM:2024a1}, where it was shown that the charge asymmetry $A_0$  can be  several percent. To quantitatively predict the magnitude of $A_0$, a model was constructed whose prediction for the contribution of $a_1(1260)$ to the cross section of the process $e^+e^-\to \pi^+\pi^-3\pi^0$ agrees with the experimental data \cite{BaBar:2018rkc}. It was also shown that due to the longitudinal polarization of the electron beam, an additional asymmetry $A_\lambda$ arises. 

In the present paper,  we study the charge asymmetry in the process $e^+e^-\to \pi^+\pi^-\pi^0\gamma$ taking into account the longitudinal polarization of electrons (positrons). We use  the model suggested in Ref.~\cite{ORM:2024a1}. The photon energy region $\nu\ll \sqrt{s}$, where $\sqrt{s}$ is the invariant mass of the $e^+e^-$ pair, is considered. In this process, it is possible to variate the  invariant mass of the virtual $a_1^0(1260)$ meson by variation of the photon energy $\nu$ (ISR method) at fixed $\sqrt{s}$ (see, i.e., the review \cite{EDSS2011}). It is shown that the electron polarization leads to a new correlation in the cross section, which provides additional opportunities for studying the process.

\section{Amplitudes of the process  $e^+e^-\to \pi^+\pi^-\pi^0\gamma$}
\begin{figure}[h!]
	\centering
	\begin{subfigure}{0.32\textwidth}
		\includegraphics[width=1\linewidth]{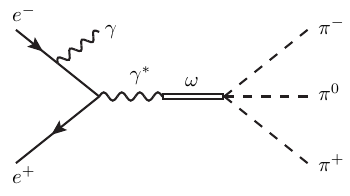}
		\caption{}
		\label{Omega_diagrams_e}
	\end{subfigure}
	\begin{subfigure}{0.32\textwidth}
		\includegraphics[width=1\linewidth]{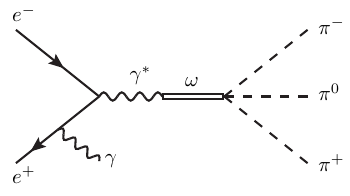}
		\caption{}
		\label{Omega_diagrams_p}
	\end{subfigure}\\
	\begin{subfigure}{0.32\textwidth}
		\includegraphics[width=1\linewidth]{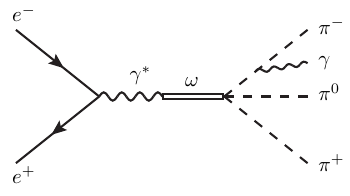}
		\caption{}
		\label{Omega_diagrams_pim}
	\end{subfigure}
	\begin{subfigure}{0.32\textwidth}
		\includegraphics[width=1\linewidth]{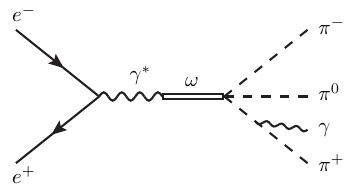}
		\caption{}
		\label{Omega_diagrams_pip}
	\end{subfigure}
	\caption{Diagrams of the process $e^+e^-\to\pi^+\pi^-\pi^0\gamma$ with an intermediate $\omega$ meson.}
	\label{Omega_diagrams}
\end{figure}

\begin{figure}[h!]
	\centering
	\begin{subfigure}{0.32\textwidth}
		\includegraphics[width=1\linewidth]{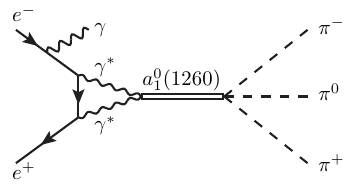}
		\caption{}
		\label{a1_diagrams_e}
	\end{subfigure}
	\begin{subfigure}{0.32\textwidth}
		\includegraphics[width=1\linewidth]{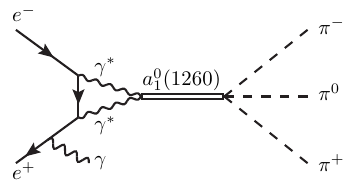}
		\caption{}
		\label{a1_diagrams_p}
	\end{subfigure}
	\begin{subfigure}{0.32\textwidth}
		\includegraphics[width=1\linewidth]{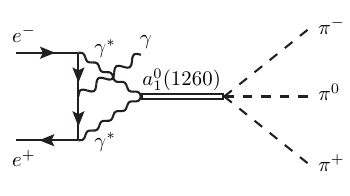}
		\caption{}
		\label{a1_diagrams_c}
	\end{subfigure}
	\\
	\begin{subfigure}{0.32\textwidth}
		\includegraphics[width=1\linewidth]{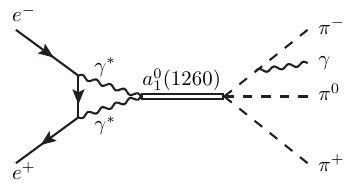}
		\caption{}
		\label{a1_diagrams_pim}
	\end{subfigure}
	\begin{subfigure}{0.32\textwidth}
		\includegraphics[width=1\linewidth]{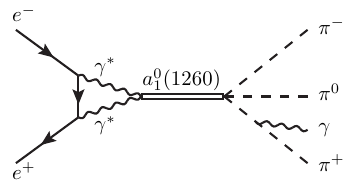}
		\caption{}
		\label{a1_diagrams_pip}
	\end{subfigure}
	\caption{Diagrams of the process $e^+e^-\to\pi^+\pi^-\pi^0\gamma$ with the intermediate $a_1^0(1260)$ meson.}
	\label{a1_diagrams}
\end{figure}

The amplitudes of the process $e^+e^-\to \pi^+\pi^-\pi^0\gamma$ in the photon energy region $\nu\ll\sqrt{s}<1.4\,\text{GeV}$ are described by the diagrams shown in Figs.~\ref{Omega_diagrams} and \ref{a1_diagrams}. In these diagrams, $\omega$ denotes a vector meson with the quantum numbers $I^G(J^{PC})=0^-(1^{--})$. For $\sqrt{s}<1.4\,\text{GeV}$, it is necessary to take into account $\omega(782)$, $\omega(1420)$, $\omega(1650)$, and $\phi(1020)$ mesons. Note that in the diagrams in Figs.~\ref{Omega_diagrams_e} and \ref{Omega_diagrams_p} the system of final pions has negative $C$ parity, while in the diagrams in 
Figs.~\ref{Omega_diagrams_pim} and \ref{Omega_diagrams_pip} it has positive $C$ parity. Thus, the main contribution to the charge asymmetry is given by the interference of diagrams in Figs.~\ref{Omega_diagrams_e} and \ref{Omega_diagrams_p} with that in Figs.~\ref{Omega_diagrams_pim} and \ref{Omega_diagrams_pip}. Therefore, in contrast to the process $e^+e^-\to\pi^+\pi^-\pi^0$, the charge asymmetry is nonzero even without accounting for the $a_1^0(1260)$ meson. Although the contribution of  $a_1^0(1260)$ meson to the charge asymmetry is not the main one, it will be shown that the contribution of $a_1^0(1260)$ meson to the charge asymmetry is not negligible.

The main contribution to the differential cross section of the process $e^+e^-\to \pi^+\pi^-\pi^0\gamma$ is given by diagrams with an intermediate $\omega$ meson. The amplitude corresponding to diagrams with the emission of a photon by the initial electron (Fig.~\ref{Omega_diagrams_e}) or positron (Fig.~\ref{Omega_diagrams_p}) has the form
\begin{align}\label{ampl_gm1}
	& \mathcal{M}_1^{(\gamma)}=\widetilde{\mathcal{M}}_1\sqrt{4\pi\alpha}\left[ \dfrac{q_1 \epsilon^*}{q_1k}-\dfrac{q_2 \epsilon^*}{q_2k} \right]\,.
\end{align}
Here $q_1$ and $q_2$ are the 4-momenta of electron and positron, $k$ and $\epsilon$ are the 4-momentum and 4-polarization vector of  photon, $\widetilde{\mathcal{M}}_1$ is the amplitude of the process $e^+e^-\to\omega\to 3\pi$ which takes into account the change in the invariant mass of $\omega$ due to photon emission (see Eq.~\eqref{ampl_mw1}). This change is important for $E=\sqrt{s}\sim m_\omega$ and $\nu\sim \Gamma_\omega$. 

The amplitudes, corresponding to the diagrams with photon emission by final $\pi^-$ (Fig.~\ref{Omega_diagrams_pim}) and $\pi^+$ (Fig.~\ref{Omega_diagrams_pip}), read
\begin{align}\label{ampl_gm2}
	& \mathcal{M}_2^{(\gamma)}=\mathcal{M}_1\sqrt{4\pi\alpha}\left[ \dfrac{p_1 \epsilon^*}{p_1k}-\dfrac{p_2 \epsilon^*}{p_2k} \right]\,,
\end{align}
where $p_1$ and $p_2$ are the 4-momenta of  $\pi^+$ and $\pi^-$, $\mathcal{M}_1$ is the amplitude of the process $e^+e^-\to\omega\to 3\pi$ (see Eq.~\eqref{ampl_m1}).

Let us discuss the diagrams with an intermediate $a_1^0(1260)$ meson (Fig.~\ref{a1_diagrams}) and consider the diagrams with photon emission in the initial state. Since the $a_1^0(1260)$ meson is produced by two virtual photons, in addition to the diagrams with photon emission by  electron (Fig.~\ref{a1_diagrams_e}) or  positron (Fig.~\ref{a1_diagrams_p}), there is the diagram (Fig.~\ref{a1_diagrams_c}) with  photon  emitted by virtual fermion.  Note that it is the sum of three diagrams that is gauge invariant. We write the amplitude of the process $e^+(q_2)e^-(q_1)\to a_1^0(q)\gamma(k)$ as
\begin{align}
	& M_{e^+e^-a\gamma}=M^{(e)}+M^{(p)}+M^{(i)}\,.
\end{align}
The amplitudes $M^{(e)}$ and  $M^{(p)}$ correspond to the diagrams with photon emission by electron and positron, respectively, $M^{(i)}$ -- by virtual fermion. These amplitudes have the form
\begin{align}\label{me}
	& M^{(e)}=\int\mbox{d}^4 k_1 T^{\alpha\beta}(k_1,k_2)(\epsilon^*)^\mu t_{\beta\alpha\mu}(k,k_2)\,,\nonumber\\
	& M^{(p)}=\int\mbox{d}^4 k_1 T^{\alpha\beta}(k_1,k_2)(\epsilon^*)^\mu t_{\mu\beta\alpha}(k_1,k)\,,\nonumber\\
	& M^{(i)}=\int\mbox{d}^4 k_1 T^{\alpha\beta}(k_1,k_2)(\epsilon^*)^\mu t_{\beta\mu\alpha}(k_1,k_2)\,,
\end{align}
where $k_2=q-k_1$, the tensors $T^{\alpha\beta}(k_1,k_2)$ and $t_{\alpha\beta\mu}(k_1,k_2)$ have the form
\begin{align}
	& T^{\alpha\beta}(k_1,k_2)=\dfrac{(4\pi\alpha)^{5/2}f_\omega f_\rho g_{a\omega\rho}\epsilon^{\mu\nu\alpha\beta} q_\mu A_\nu^*}{(2\pi)^4 k_1^2 k_2^2}\left[ \dfrac{1}{\mathcal{D}_\omega(k_1)\mathcal{D}_\rho(k_2)}-\dfrac{1}{\mathcal{D}_\omega(k_2)\mathcal{D}_\rho(k_1)} \right]\,,\nonumber\\
	& t_{\alpha\beta\mu}(k_1,k_2)=\dfrac{\bar{V}\gamma_\alpha(-\hat{q}_2+\hat{k}_2+m_e)\gamma_\beta(\hat{q}_1-\hat{k}_1+m_e)\gamma_\mu U}{[(q_1-k_1)^2-m_e^2][(-q_2+k_2)^2-m_e^2]}\,,\nonumber\\
	& \mathcal{D}_\omega(q)=q^2-m_\omega^2+i\Gamma_\omega m_\omega\,,\quad \mathcal{D}_\rho(q)=q^2-m_\rho^2+i\Gamma_\rho m_\rho\,,\nonumber\\
	& q=q_1+q_2-k\,,\quad \epsilon^{0123}=1\,.
\end{align}
Here $V$ and $U$ are bispinors corresponding to the negative and positive energy solutions of the Dirac equation, $A_\mu$ is the polarization vector of the $a_1^0(1260)$ meson, $f_\omega$ and $f_\rho$ are the coupling constants of  $\omega$ and $\rho$ mesons with the photon, $g_{a\omega\rho}$ is the coupling constant at the vertex $a_1(1260)\omega\rho$, $m_e$ is the electron mass, $m_\omega$ and $\Gamma_\omega$ are the mass and width of the $\omega$ meson, $m_\rho$ and $\Gamma_\rho$ are the mass and width of the $\rho$ meson. Then, using the Sudakov method \cite{Sudakov:1954sw}, we take the integrals in the amplitudes \eqref{me} with the logarithmic accuracy in the parameter $E/m_e$. For  electron with the helicity $\lambda$, positron with the helicity $-\lambda$, and the photon energy $\nu\ll m_e$, we have
 \begin{align}
	& M^{(e)}=2C_a\ln{\left( \dfrac{q^2}{m_e^2} \right)}\left( \dfrac{q_1\epsilon^*}{q_1k} \right)\,,\quad M^{(p)}=-2C_a\ln{\left( \dfrac{q^2}{m_e^2} \right)}\left( \dfrac{q_2\epsilon^*}{q_2k} \right)\,,\quad M^{(i)}=0\,,\nonumber\\
	& C_a=\lambda\,(\bm e_\lambda\cdot\bm A)\,\dfrac{(4\pi\alpha)^{5/2}f_\omega f_\rho g_{a\omega\rho} }{(4\pi)^2 q^2\mu_\omega^2\mu_\rho^2}\left\langle \mu_\omega^2\ln{\left( 1-\dfrac{q^2}{\mu_\omega^2} \right)}-\mu_\rho^2\ln{\left( 1-\dfrac{q^2}{\mu_\rho^2} \right)} \right\rangle\,,\nonumber\\
	& \bm e_\lambda=\bm e_x+i\lambda\bm e_y\,,\quad \mu_\omega^2=m_\omega^2-i\Gamma_\omega m_\omega\,,\quad \mu_\rho^2=m_\rho^2-i\Gamma_\rho m_\rho\,,
\end{align}
where $\lambda=\pm 1$, $\bm e_x$ and $\bm e_y$ are two unit vectors orthogonal to the electron momentum.

In another limiting case $m_e\ll\nu\ll E$ we find
\begin{align}
	M^{(e)}&=C_a\left[ \ln{\left( \dfrac{q^2}{m_e^2} \right)}+\ln{\left( \dfrac{q^2}{q_1k} \right)} \right]\left( \dfrac{q_1\epsilon^*}{q_1k} \right)\,,\nonumber\\
	M^{(p)}&=-C_a\left[ \ln{\left( \dfrac{q^2}{m_e^2} \right)}+\ln{\left( \dfrac{q^2}{q_2k} \right)} \right]\left( \dfrac{q_2\epsilon^*}{q_2k} \right)\,,\nonumber\\
	M^{(i)}&=C_a\left\langle \left[ \ln{\left( \dfrac{q^2}{m_e^2} \right)}-\ln{\left( \dfrac{q^2}{q_1k} \right)} \right]\left( \dfrac{q_1\epsilon^*}{q_1k} \right)\right.\nonumber\\
	& \left.-\left[ \ln{\left( \dfrac{q^2}{m_e^2} \right)}-\ln{\left( \dfrac{q^2}{q_2k} \right)} \right]\left( \dfrac{q_2\epsilon^*}{q_2k} \right)  \right\rangle\,.
\end{align}
As a result, for the amplitudes $M^{(e)}$, $M^{(p)}$ and $M^{(i)}$, we obtain with the logarithmic accuracy the expressions valid in all region  $\nu\ll E$, 
\begin{align}
	M^{(e)}&=C_a\left[ \ln{\left( \dfrac{q^2}{m_e^2} \right)}+\ln{\left( \dfrac{q^2}{m_e^2+q_1k} \right)} \right]\left( \dfrac{q_1\epsilon^*}{q_1k} \right)\,,\nonumber\\
	M^{(p)}&=-C_a\left[ \ln{\left( \dfrac{q^2}{m_e^2} \right)}+\ln{\left( \dfrac{q^2}{m_e^2+q_2k} \right)} \right]\left( \dfrac{q_2\epsilon^*}{q_2k} \right)\,,\nonumber\\
	M^{(i)}&=C_a\left\langle \left[ \ln{\left( \dfrac{q^2}{m_e^2} \right)}-\ln{\left( \dfrac{q^2}{m_e^2+q_1k} \right)} \right]\left( \dfrac{q_1\epsilon^*}{q_1k} \right)\right.\nonumber\\
	& \left.-\left[ \ln{\left( \dfrac{q^2}{m_e^2} \right)}-\ln{\left( \dfrac{q^2}{m_e^2+q_2k} \right)} \right]\left( \dfrac{q_2\epsilon^*}{q_2k} \right)  \right\rangle\,.
\end{align}
Numerically these formulas  are in good agreement with the results of calculations performed with the use of PackageX \cite{Patel:2015tea,Patel:2016fam}.
Summing the amplitudes $M^{(e)}$, $M^{(p)}$ and $M^{(i)}$, we find
\begin{align}\label{meeag}
	& M_{e^+e^-a\gamma}=2C_a\ln{\left( \dfrac{q^2}{m_e^2} \right)}\left[ \dfrac{q_1\epsilon^*}{q_1k}-\dfrac{q_2\epsilon^*}{q_2k} \right]\,.
\end{align}
Thus, the amplitude of photon emission with $\nu\ll E$ in the case of meson production via two virtual photons is similar to that of meson production via one virtual photon. We emphasize that in the first case the diagram with photon emission by virtual fermion (Fig.~\ref{a1_diagrams_c}) should be taken into account. Otherwise the result would be  incorrect for any $\nu\ll E$.

Then, summing the  amplitudes for the diagrams in Figs.~\ref{a1_diagrams_e}, \ref{a1_diagrams_p} and \ref{a1_diagrams_c}, we obtain
\begin{align}\label{ampl_gm3}
	& \mathcal{M}_3^{(\gamma)}=\widetilde{\mathcal{M}}_2\sqrt{4\pi\alpha}\left[ \dfrac{q_1 \epsilon^*}{q_1k}-\dfrac{q_2 \epsilon^*}{q_2k} \right]\,,
\end{align}
where $\widetilde{\mathcal{M}}_2$ is the amplitude of the process $e^+e^-\to a_1^0(1260)\to 3\pi$, which  takes into account the change of $a_1^0(1260)$ invariant mass due to  photon emission (see Eq.~\eqref{ampl_mw2}). Then,  the amplitude corresponding to the diagrams with photon emission by final $\pi^-$ (Fig.~\ref{a1_diagrams_pim}) or $\pi^+$ (Fig.~\ref{a1_diagrams_pip}) reads
\begin{align}\label{ampl_gm4}
	& \mathcal{M}_4^{(\gamma)}=\mathcal{M}_2\sqrt{4\pi\alpha}\left[ \dfrac{p_1 \epsilon^*}{p_1k}-\dfrac{p_2 \epsilon^*}{p_2k} \right]\,.
\end{align}
Here $\mathcal{M}_2$ is the amplitude of the process $e^+e^-\to a_1^0(1260)\to 3\pi$ (see Eq.~\eqref{ampl_m2}).

\section{Cross section of the process $e^+e^-\to \pi^+\pi^-\pi^0\gamma$}
 In the photon energy region $\nu\ll E$,  the differential in the  momenta of charged pions and the photon angles $\mbox{d}\Omega_{\bm k}$ cross section, $\mbox{d}\sigma_\gamma(\bm p_1,\bm p_2,\bm k)$,  for a polarized electron and unpolarized positron has the form
\begin{align}\label{idifg}
	& \mbox{d}\sigma_\gamma(\bm p_1,\bm p_2,\bm k)=\sum_{pol}\dfrac{|\mathcal{M}^{(\gamma)}|^2}{32(2\pi)^8}|\bm p_1||\bm p_2|\delta(B)\mbox{d}\Omega_1\mbox{d}\Omega_2\mbox{d}\varepsilon_1\mbox{d}\varepsilon_2\,\nu\mbox{d}\nu\mbox{d}\Omega_{\bm k}\,, \nonumber\\
	&\mathcal{M}^{(\gamma)}=\mathcal{M}^{(\gamma)}_1+\mathcal{M}^{(\gamma)}_2+\mathcal{M}^{(\gamma)}_3+\mathcal{M}^{(\gamma)}_4\,,\quad B=(q_1+q_2-p_1-p_2)^2-m_\pi^2\,,
\end{align}
where $\mathcal{M}^{(\gamma)}_1$, $\mathcal{M}^{(\gamma)}_2$, $\mathcal{M}^{(\gamma)}_3$, and $\mathcal{M}^{(\gamma)}_4$ are given in Eqs.~\eqref{ampl_gm1}, \eqref{ampl_gm2}, \eqref{ampl_gm3}, and \eqref{ampl_gm4},  the summation is performed over the photon polarizations, $m_\pi$ is the pion mass. Note that the  delta function in Eq.~\eqref{idifg}  can be eliminated by integration  either over the angles or over the energies of $\pi^+$ and $\pi^-$. The cross section $\mbox{d}\sigma_\gamma(\bm p_1,\bm p_2,\bm k)$ can be conveniently represented as
\begin{align}
	& \mbox{d}\sigma_\gamma(\bm p_1,\bm p_2,\bm k)=\mbox{d}\sigma_{12}(\bm p_1,\bm p_2,\bm k)+\mbox{d}\sigma_{13}(\bm p_1,\bm p_2,\bm k)+\mbox{d}\sigma_{24}(\bm p_1,\bm p_2,\bm k)\,,\nonumber\\
	& \mbox{d}\sigma_a(\bm p_1,\bm p_2,\bm k)=\sum_{pol}\dfrac{\mathcal{N}_a}{32(2\pi)^8}|\bm p_1||\bm p_2|\delta(B)\mbox{d}\Omega_1\mbox{d}\Omega_2\mbox{d}\varepsilon_1\mbox{d}\varepsilon_2\,\nu\mbox{d}\nu\mbox{d}\Omega_{\bm k}\,,\quad a=\{ 12\},\,\{13\},\,\{24 \}\,,\nonumber\\
	& \mathcal{N}_{12}=2\mbox{Re}\left[ \left( \mathcal{M}^{(\gamma)}_1+\mathcal{M}^{(\gamma)}_3 \right) \left( \mathcal{M}^{(\gamma)}_2+\mathcal{M}^{(\gamma)}_4 \right)^* \right]\approx 2\mbox{Re}\left[ \mathcal{M}^{(\gamma)}_1 \mathcal{M}^{(\gamma)*}_2 \right]\,,\nonumber\\
	& \mathcal{N}_{13}=\left| \mathcal{M}^{(\gamma)}_1+\mathcal{M}^{(\gamma)}_3 \right|^2\approx \left| \mathcal{M}^{(\gamma)}_1 \right|^2+2\mbox{Re}\left[ \mathcal{M}^{(\gamma)}_1 \mathcal{M}^{(\gamma)*}_3 \right] \,,\nonumber\\
	& \mathcal{N}_{24}=\left| \mathcal{M}^{(\gamma)}_2+\mathcal{M}^{(\gamma)}_4 \right|^2\approx \left| \mathcal{M}^{(\gamma)}_2 \right|^2\,.
\end{align}
We keep only the main contribution to $\mathcal{N}_{12}$, $\mathcal{N}_{13}$, and $\mathcal{N}_{24}$. 

In our work we consider an experiment in which a photon with the energy $\nu$ is detected. Therefore, we  introduce the cross section
\begin{align}
\mbox{d}\sigma_\gamma(\bm p_1,\bm p_2)=\int \nu\,\dfrac{\mbox{d}\sigma_\gamma(\bm p_1,\bm p_2,\bm k)}{\mbox{d}\nu\mbox{d}\Omega_{\bm k}}\mbox{d}\Omega_{\bm k}\,,
\end{align}
which is differential in $\nu$ and integral in the photon emission angles. We represent the cross section $\mbox{d}\sigma_\gamma(\bm p_1,\bm p_2)$ as the sum of three terms,
\begin{align}\label{dsig_g}
	& \mbox{d}\sigma_\gamma(\bm p_1,\bm p_2)=\mbox{d}\sigma_{12}(\bm p_1,\bm p_2)+\mbox{d}\sigma_{13}(\bm p_1,\bm p_2)+\mbox{d}\sigma_{24}(\bm p_1,\bm p_2)\,.
\end{align}
Where $\mbox{d}\sigma_{12}(\bm p_1,\bm p_2)$ is given by  the term  $\propto\mathcal{N}_{12}$,
\begin{align}\label{dsig_12}
	& \mbox{d}\sigma_{12}(\bm p_1,\bm p_2)=\alpha\ln{\left[ \dfrac{(\varepsilon_1+\bm p_1\cdot\bm N)(\varepsilon_2-\bm p_2\cdot\bm N)}{(\varepsilon_1-\bm p_1\cdot\bm N)(\varepsilon_2+\bm p_2\cdot\bm N)} \right]}\nonumber\\
&\times	\dfrac{\mbox{Re}[\widetilde{\mathcal{M}}_1 \mathcal{M}_1^*]}{4(2\pi)^6}|\bm p_1||\bm p_2|\delta(B)\mbox{d}\Omega_1\mbox{d}\Omega_2\mbox{d}\varepsilon_1\mbox{d}\varepsilon_2\,.
\end{align}
Here $\bm N$ is a unit vector directed along the electron momentum.
The term  $\propto\mathcal{N}_{13}$ contributes to $\mbox{d}\sigma_{13}(\bm p_1,\bm p_2)$,
\begin{align}
	& \mbox{d}\sigma_{13}(\bm p_1,\bm p_2)=\alpha\ln{\left( \dfrac{E^2}{m_e^2} \right)}\dfrac{|\widetilde{\mathcal{M}}_1|^2+2\mbox{Re}({\widetilde{\mathcal{M}}_1}^* \widetilde{\mathcal{M}}_2)}{4(2\pi)^6}|\bm p_1||\bm p_2|\delta(B)\mbox{d}\Omega_1\mbox{d}\Omega_2\mbox{d}\varepsilon_1\mbox{d}\varepsilon_2\,.\label{dsig_13}
\end{align}
Finally, the term $\propto\mathcal{N}_{24}$ gives the  contribution to $\mbox{d}\sigma_{24}(\bm p_1,\bm p_2)$, 
\begin{align}
	&\mbox{d}\sigma_{24}(\bm p_1,\bm p_2)=\alpha\left\langle -1+\dfrac{\eta}{\sqrt{\eta^2-m_\pi^4}}\ln{\left( \dfrac{\eta+\sqrt{\eta^2-m_\pi^4}}{m_\pi^2} \right)} \right\rangle\nonumber\\
	& \times \dfrac{|\mathcal{M}_1|^2+2\mbox{Re}(\mathcal{M}_1^*\mathcal{M}_2)}{4(2\pi)^6}|\bm p_1||\bm p_2|\delta(B)\mbox{d}\Omega_1\mbox{d}\Omega_2\mbox{d}\varepsilon_1\mbox{d}\varepsilon_2\,,\label{dsig_24}
	\nonumber\\
	&\eta=p_1p_2=\varepsilon_1\varepsilon_2-\bm p_1\bm p_2> m_\pi^2\,.
\end{align}
In Eqs.~\eqref{dsig_12}, \eqref{dsig_13}, and \eqref{dsig_24}  a smallness of $m_e/E$  is used.

\begin{figure}[h!] 
	\centering
	\includegraphics[width=0.49\linewidth]{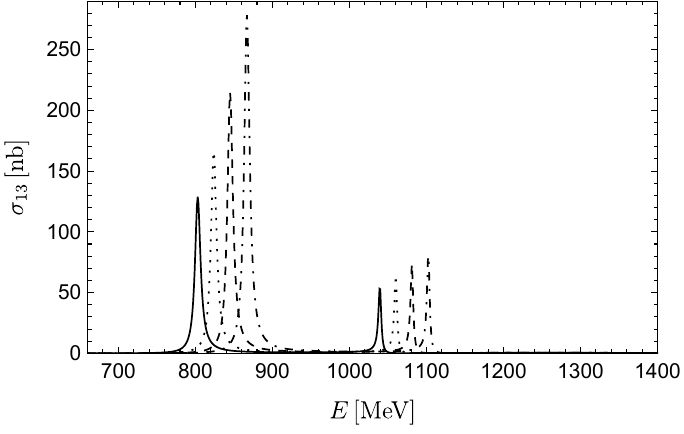}
	\includegraphics[width=0.49\linewidth]{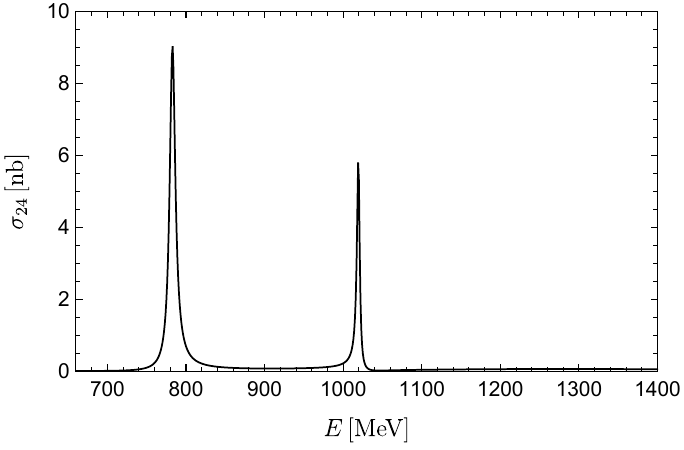}
	\caption{Left plot: the energy dependence of $\sigma_{13}$ for $\nu=20\,\mbox{MeV}$ (solid curve),  $\nu=40\,\mbox{MeV}$ (dotted curve), $\nu=60\,\mbox{MeV}$  (dashed curve), and $\nu=80\,\mbox{MeV}$ (dash-dotted curve). Right plot: the energy dependence of $\sigma_{24}$.}
	\label{sig13_24}
\end{figure}

Integrating $\mbox{d}\sigma_\gamma(\bm p_1,\bm p_2)$ over the energies and angles of pions, we obtain the total cross section 
$$\sigma_\gamma=\sigma_{12}+\sigma_{13}+\sigma_{24}\,.$$
 The term $\sigma_{12}$ is given by interference of the diagrams with photon emission from the initial and final states, $\sigma_{13}$ is given by the diagrams with photon emission from the initial state, and $\sigma_{24}$ is given by photon emission from the final state. The main contribution to the cross section $\sigma_\gamma$ is given by $\sigma_{13}$, and the contribution of $\sigma_{12}$ can be neglected. In the left plot of Fig.~\ref{sig13_24}, the energy dependence of $\sigma_{13}$ is shown for a few values of $\nu$. 
It is seen that there are two peaks in the region of $\omega(782)$ and $\phi(1020)$ masses. Moreover, the peaks grow and shift to the right with increasing photon energy. The energy  dependence of $\sigma_{24}$ is shown in the right plot of Fig.~\ref{sig13_24}. Unlike $\sigma_{13}$, the term $\sigma_{24}$ is independent of $\nu$ at $\nu\ll E$.  Note that $\sigma_{24}$ has also two peaks in the region of masses $\omega(782)$ and $\phi(1020)$, but the magnitudes of these peaks are essentially smaller than that for $\sigma_{13}$.

\section{Charge asymmetry in the process $e^+e^-\to \pi^+\pi^-\pi^0\gamma$}
Similarly to Ref.~\cite{ORM:2024a1}, we  define the differential charge asymmetry as
\begin{align}
	& \mbox{d}A_\gamma(\bm p_1,\bm p_2)=\dfrac{\mbox{d}\sigma_\gamma(\bm p_1,\bm p_2)-\mbox{d}\sigma_\gamma(\bm p_2,\bm p_1)}{2\sigma_\gamma}\,.
\end{align}
Note that  this expression has no singularities at $\nu=0$. We  represent the differential asymmetry as the sum of three terms,
\begin{align}
	& \mbox{d}A_\gamma(\bm p_1,\bm p_2)=\mbox{d}A_{12}(\bm p_1,\bm p_2)+\mbox{d}A_{13}(\bm p_1,\bm p_2)+\mbox{d}A_{24}(\bm p_1,\bm p_2)\,,\nonumber\\
	& \mbox{d}A_a(\bm p_1,\bm p_2)=\dfrac{\mbox{d}\sigma_a(\bm p_1,\bm p_2)-\mbox{d}\sigma_a(\bm p_2,\bm p_1)}{2\sigma_\gamma}\,,\quad a=\{ 12\},\,\{13\},\,\{24 \}\,.
\end{align}
Each of the terms, integrated over the angles and energies of  pions, vanishes. Therefore, it is necessary to integrate over a part of the phase space.

Let us introduce the integral asymmetry $A^{\gamma}_0$ independent of the electron helicity $\lambda$,
\begin{align}\label{A0}
	& A^{\gamma}_0=\int\Xi_0(\bm p_1,\bm p_2)\mbox{d}A_\gamma(\bm p_1,\bm p_2)=A^{(12)}_0+A^{(13)}_0+A^{(24)}_0\,,\nonumber\\
	& A^{(a)}_0=\int\Xi_0(\bm p_1,\bm p_2)\mbox{d}A_a(\bm p_1,\bm p_2)\,,\quad a=\{ 12\},\{13\},\{24 \}\,,\nonumber\\
	& \Xi_0(\bm p_1,\bm p_2)=\theta(\bm N\cdot\bm p_1)\theta(-\bm N\cdot\bm p_2)-\theta(-\bm N\cdot\bm p_1)\theta(\bm N\cdot\bm p_2)\,.
\end{align}
Here $\theta(x)$ is the Heaviside theta function, and the function $\Xi_0(\bm p_1,\bm p_2)$ selects a part of phase space in which $\pi^+$ flies to the forward hemisphere, and $\pi^-$ -- to the backward one, or vice versa.
In Eq.~\eqref{A0}, the term $A_0^{(12)}$ arises as a result of interference of the diagrams with photon emission of initial and final states, $A_0^{(13)}$ arises as a result of interference of the diagrams with photon  emission only of the initial state, and $A_0^{(24)}$ arises as a result of interference of the diagrams with emission only of the final state. The main contribution to the asymmetry $A_0^\gamma$ is given by $A_0^{(12)}$, and the contribution of $A_0^{(24)}$ can be neglected. The diagram with $a_1^0(1260)$ meson in the intermediate state does not contribute to  $A_0^{(12)}$. Thus, in contrast to the process $e^+e^-\to \pi^+\pi^-\pi^0$, the charge asymmetry is nonzero even without accounting for the contribution of $a_1^0(1260)$ meson. 

\begin{figure}[h!] 
	\centering
	\includegraphics[width=0.49\linewidth]{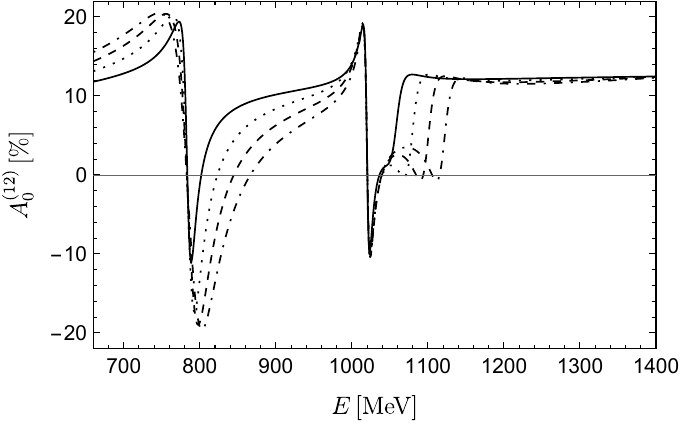}
	\includegraphics[width=0.49\linewidth]{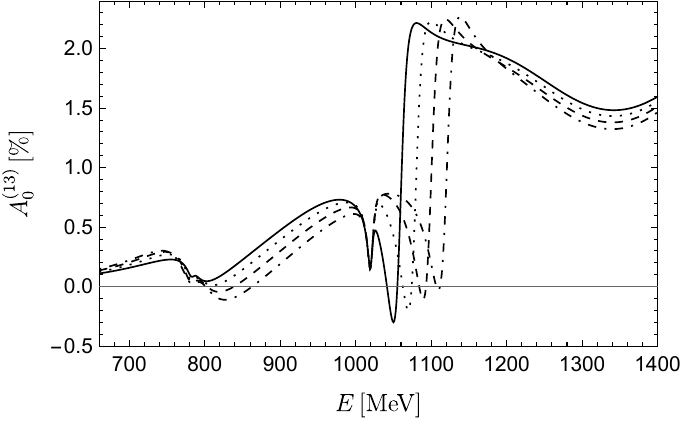}
	\caption{Dependence of $A_0^{(12)}$ (left plot) and  $A_0^{(13)}$ (right plot) on the energy $E$  for $\nu=20\,\mbox{MeV}$ (solid curve), $\nu=40\,\mbox{MeV}$ (dotted curve), $\nu=60\,\mbox{MeV}$ (dashed curve), and $\nu=80\,\mbox{MeV}$ (dash-dotted curve).}
	\label{A0_12_13}
\end{figure}

In  Fig.~\ref{A0_12_13}, the energy dependence of $A_0^{(12)}$ (left plot) and  $A_0^{(13)}$ (right plot) is shown  for $\nu=20\,\mbox{MeV}$ (solid curve), $\nu=40\,\mbox{MeV}$ (dotted curve), $\nu=60\,\mbox{MeV}$ (dashed curve), and $\nu=80\,\mbox{MeV}$ (dash-dotted curve).
Although $A_0^{(12)}$  can reach $20\%$,  the total cross section $\sigma_\gamma$ in the corresponding energy region is  quite small (see Fig.~\ref{sig13_24}). Therefore, it is easier to observe the asymmetry near the peaks, where the energy dependence of the asymmetry  is very strong. There are peaks in $A_0^{(12)}$ related to $\omega$ mesons in intermediate state  in the amplitudes $\mathcal{M}_1^{(\gamma)}$ and $\mathcal{M}_2^{(\gamma)}$. The magnitude of asymmetry in the peaks increases with increasing photon energy, and the position of peaks shifts toward higher $E$.

Further, the term $A_0^{(13)}$ is approximately an order of magnitude smaller than  $A_0^{(12)}$. In addition, we recall that the relative phase between $\omega(782)$ and $a_1(1260)$ is unknown. Therefore, for the sake of example,  we set it equal to zero as in Ref.~\cite{ORM:2024a1}. 

\begin{figure}[h!] 
	\centering
	\includegraphics[width=0.49\linewidth]{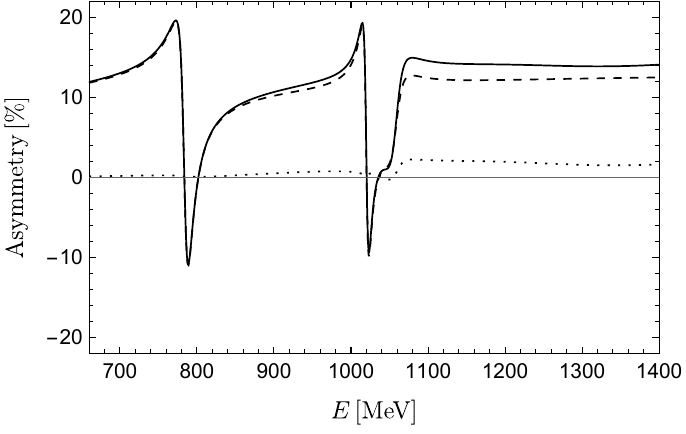}
	\includegraphics[width=0.49\linewidth]{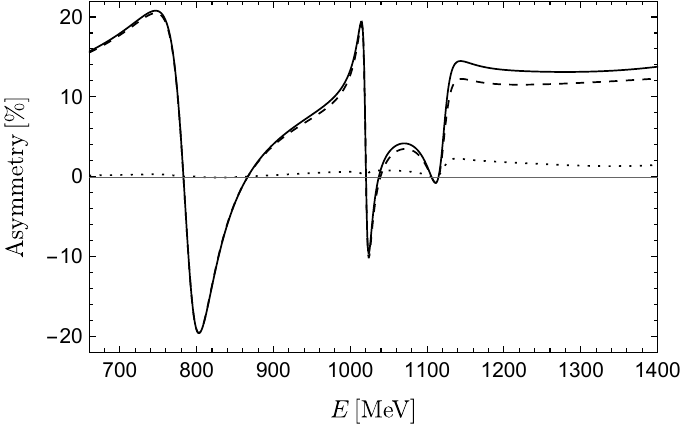}
	\caption{Energy dependence of $A_0^\gamma$ (solid curve), $A_0^{(12)}$ (dashed curve), and $A_0^{(13)}$ (dotted curve) for $\nu=20\,\mbox{MeV}$ (left plot) and $\nu=80\,\mbox{MeV}$ (right plot).}
	\label{A_Comp}
\end{figure}

Although the contribution of $A_0^{(13)}$ is substantially smaller than that of $A_0^{(12)}$, it is nevertheless noticeable, see Fig.~\ref{A_Comp}.
It is seen that the contribution of $A_0^{(13)}$ affects the height of the plateau  at energies  $E\gtrsim 1100\,\mbox{MeV}$.  Since $A_0^{(13)}$ is nonzero due to accounting for $a_1^0(1260)$ meson, observing its manifestation in the charge asymmetry is not a hopeless task.

Let us now introduce the integral asymmetry $A^{\gamma}_\lambda$ proportional to the electron helicity $\lambda$,
\begin{align}
&	A^{\gamma}_\lambda=\int\Xi_\lambda(\bm p_1,\bm p_2)\mbox{d}A_\gamma(\bm p_1,\bm p_2)=A^{(12)}_\lambda+A^{(13)}_\lambda+A^{(24)}_\lambda\,,\nonumber\\
&	A^{(a)}_\lambda=\int\Xi_\lambda(\bm p_1,\bm p_2)\mbox{d}A_a(\bm p_1,\bm p_2)\,,\quad a=\{ 12\},\,\{13\},\,\{24 \}\,,\nonumber\\
&	\Xi_\lambda(\bm p_1,\bm p_2)=\left\{ \theta(\bm N\cdot[\bm p_1\times\bm p_2])-\theta(-\bm N\cdot[\bm p_1\times\bm p_2]) \right\}\nonumber\\
	& \times\left\{ \theta(\bm N\cdot\bm p_1)\theta(\bm N\cdot\bm p_2)-\theta(-\bm N\cdot\bm p_1)\theta(-\bm N\cdot\bm p_2) \right\}\,,
\end{align}
where the function $\Xi_\lambda(\bm p_1,\bm p_2)$ selects the region of phase space in which both pions fly either to the forward or to the backward hemisphere. The sign of $\Xi_\lambda(\bm p_1,\bm p_2)$ is determined by the sign of the correlation $(\bm N\cdot[\bm p_1\times\bm p_2])$. The term $A_\lambda^{(12)}$ arises due to interference of the diagrams with photon emission of the initial and final states, $A_\lambda^{(13)}$ -- due to interference of the diagrams with photon emission only of the initial state, and $A_\lambda^{(24)}$ -- due to interference of the diagrams with photon emission only of the final state. It turned out that the term $A_\lambda^{(12)}$ vanishes, and the term $A_\lambda^{(24)}$ can be neglected compared to $A_\lambda^{(13)}$. Therefore,  $A_\lambda^\gamma\approx A_\lambda^{(13)}$ and  the asymmetry $A_\lambda^{\gamma}$ differs from zero only due to  accounting for $a_1^0(1260)$ meson in the intermediate state.

\begin{figure}[h!] 
	\centering
	\includegraphics[width=0.6\linewidth]{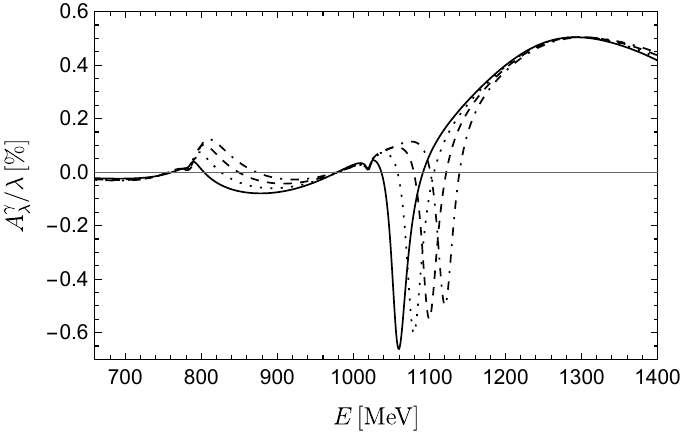}
	\caption{Energy dependence of $A_\lambda^{\gamma}/\lambda$  for $\nu=20\,\mbox{MeV}$ (solid curve), $\nu=40\,\mbox{MeV}$ (dotted curve), $\nu=60\,\mbox{MeV}$ (dashed curve), and $\nu=80\,\mbox{MeV}$ (dash-dotted curve).
	}
	\label{Alg}
\end{figure}

 The energy dependence of $A_\lambda^{\gamma}/\lambda$ is shown in Fig.~\ref{Alg} for a few values of $\nu$. 
It is seen that the positions of minimum and maximum shift towards larger $E$ with increasing photon energy.

\section{Conclusion}
In our work, the charge asymmetry in the process $e^+e^-\to \pi^+\pi^-\pi^0\gamma$ is investigated with account for the longitudinal polarization of electrons (positrons). The asymmetry arises due to interference of the amplitudes with different $C$ parities  of the pion system.
The integral asymmetries $A_0^\gamma$ and $A_\lambda^\gamma$, independent of polarization and dependent on it,  are analyzed. The  contribution of $a_1^0(1260)$ meson to $A_0^\gamma$  is taken into account. Although the asymmetry $A_0^\gamma$ is nonzero even without accounting for $a_1^0(1260)$ meson, its manifestation in the charge asymmetry is not negligible. It is shown that the charge asymmetry $A_0^\gamma$ can reach several tens of percent.  The asymmetry $A_\lambda^\gamma$ is essentially smaller than $A_0^\gamma$, but is nonzero only with accounting for $a_1^0(1260)$ meson in the  intermediate state. 

\appendix 
\section{Amplitudes for $e^+e^-\to\omega\to 3\pi$ and $e^+e^-\to a_1\to 3\pi$ processes
}
The amplitude of the process $e^+(q_2)e^-(q_1)\to\omega\to \pi^+(p_1)\pi^-(p_2)\pi^0(p_3)$ reads
\begin{align}\label{ampl_m1}
	& \mathcal{M}_1=\dfrac{4\pi\alpha f_\omega F_\omega}{Q^2\mathcal{D}_\omega(Q)}\bm e_\lambda\cdot[\bm p_1\times\bm p_2]\,,\quad Q^2=(q_1+q_2)^2=E^2\,,\nonumber\\
	& F_\omega=\dfrac{2g_{\rho\pi\pi}g_{\omega\rho\pi} E}{m_\omega}\left\langle \dfrac{1}{\mathcal{D}_\rho(p_1+p_2)}+\dfrac{1}{\mathcal{D}_\rho(p_1+p_3)}+\dfrac{1}{\mathcal{D}_\rho(p_2+p_3)}\right\rangle\,,\nonumber\\
	& \bm e_\lambda=\bm e_x+i\lambda\bm e_y\,,\quad \mathcal{D}_\omega(q)=q^2-m_\omega^2+i\Gamma_\omega m_\omega\,,\quad \mathcal{D}_\rho(q)=q^2-m_\rho^2+i\Gamma_\rho m_\rho\,,
\end{align}
where $\alpha$ is the fine structure constant, $f_\omega$ is the coupling constant of  $\omega$ meson and photon, $g_{\rho\pi\pi}$ is the coupling constant at $\rho\pi\pi$ vertex, $g_{\omega\rho\pi}$ is the coupling constant at  $\omega\rho\pi$ vertex, $m_\omega$ and $\Gamma_\omega$ are the mass and width of $\omega$ meson, $m_\rho$ and $\Gamma_\rho$ are the mass and width of  $\rho$ meson, $\lambda=\pm 1$ is the electron helicity, $\bm e_x$ and $\bm e_y$ are two unit vectors orthogonal to the electron momentum. 

The amplitude of the process $e^+(q_2)e^-(q_1)\to a_1^0(1260)\to \pi^+(p_1)\pi^-(p_2)\pi^0(p_3)$ is
\begin{align}\label{ampl_m2}
	& \mathcal{M}_2=\lambda\bm e_\lambda\cdot\left[ \bm p_1 Z(\varepsilon_1,\varepsilon_2)+\bm p_2 Z(\varepsilon_2,\varepsilon_1) \right]F_a\,,\nonumber\\
	& F_a=-i\dfrac{2E\alpha^2 g_{\rho\pi\pi}g_{a\rho\pi}g_{a\omega\rho}}{m_a\mathcal{D}_a(Q)}\mathcal{G}_a(E)\,,\quad Z(\varepsilon_1,\varepsilon_2)=\dfrac{E-\varepsilon_2}{\mathcal{D}_\rho(p_1+p_3)}+\dfrac{\varepsilon_2}{\mathcal{D}_\rho(p_2+p_3)}\,,\nonumber\\
	& \mathcal{G}_a(E)=\dfrac{4f_\omega f_\rho}{E^2\mu_\omega^2\mu_\rho^2}\ln{\left( \dfrac{E}{m_e} \right)}\left[ \mu_\omega^2\ln{\left( 1-\dfrac{E^2}{\mu_\omega^2} \right)}-\mu_\rho^2\ln{\left( 1-\dfrac{E^2}{\mu_\rho^2} \right)} \right]\,,\nonumber\\
	& \mathcal{D}_a(q)=q^2-m_a^2+i\Gamma_a m_a\,,\quad \mu_\omega^2=m_\omega^2-i\Gamma_\omega m_\omega\,,\quad \mu_\rho^2=m_\rho^2-i\Gamma_\rho m_\rho\,.
\end{align}
Here $f_\rho$ is the coupling constant of  $\rho$ meson and photon, $g_{a\rho\pi}$ is the coupling constant at the vertex $a_1(1260)\rho\pi$, $g_{a\omega\rho}$ is the coupling constant at the vertex $a_1(1260)\omega\rho$, $m_a$ and $\Gamma_a$ are the mass and width of  $a_1(1260)$ meson, $m_e$ is the electron mass.
For more details see Ref.~\cite{ORM:2024a1}.

The amplitude $\widetilde{\mathcal{M}}_1$,  contributing to Eq.~\eqref{ampl_gm1}, has the form 
\begin{align}
	&\widetilde{\mathcal{M}}_1 =\dfrac{4\pi\alpha f_\omega F_\omega}{Q^2\mathcal{D}_\omega(Q-k)}\bm e_\lambda\cdot[\bm p_1\times\bm p_2]\label{ampl_mw1}\,.
\end{align}
Unlike $\mathcal{M}_1$, this amplitude takes into account the change  of  $\omega$ meson  invariant mass due to photon emission. This effect is important for $E\sim m_\omega$ and $\nu\sim \Gamma_\omega$. Also, the amplitude $\widetilde{\mathcal{M}}_2$,  contributing to Eq.~\eqref{ampl_gm3}, reads
\begin{align}
	& \widetilde{\mathcal{M}}_2=\lambda\bm e_\lambda\cdot\left[ \bm p_1 Z(\varepsilon_1,\varepsilon_2)+\bm p_2 Z(\varepsilon_2,\varepsilon_1) \right]\widetilde{F}_a\label{ampl_mw2}\,,\nonumber\\
	& \widetilde{F}_a=-i\dfrac{2E\alpha^2 g_{\rho\pi\pi}g_{a\rho\pi}g_{a\omega\rho}}{m_a\mathcal{D}_a(Q-k)}\mathcal{G}_a(\sqrt{E(E-2\nu)})\,.
\end{align}
This amplitude takes into account the change of  $a_1^0(1260)$ meson invariant mass due to photon emission. For $\nu\ll E$ and $\nu\ll \Gamma_a$,  a variation  of  $a_1^0(1260)$ meson invariant mass can be neglected. Thus, for the region of $\nu$  considered in our work, the shift in $\widetilde{\mathcal{M}}_1$ should be taken into account only for $\omega(782)$ and $\phi(1020)$, while the shift in $\widetilde{\mathcal{M}}_2$  is unimportant.

\section*{Acknowledgement}
The authors thank R.N. Lee for useful discussions. The work of I.V. Obraztsov was supported by the Foundation for the Advancement of Theoretical Physics and Mathematics ”BASIS” under Grant No. 24-1-5-3-1.

\end{document}